\documentclass[a4paper,twocolumn,prl]{revtex4-1}
\usepackage{graphicx}
\usepackage{amsmath}           
\usepackage{amsfonts}          
\usepackage{epstopdf}
\usepackage{amsthm}

\usepackage{newtxtext,newtxmath}

\usepackage{tikz,pgfplots,bbm,bm,mathrsfs,dsfont}
\usepackage{subcaption}
\usepackage{color}
\usepackage{quoting}

\usepackage[english]{babel}
\usepackage{hyperref}
\hypersetup{pdfauthor={Capelli Caracciolo DiGioacchino Malatesta},pdftitle={Exact bound for a TSP cost in two dimensions},%
	colorlinks, linktocpage=true, pdfstartpage=3, pdfstartview=FitV,%
	urlcolor=orange, linkcolor=blue, citecolor=red, 
}

\newcommand{\be}{\begin{equation}}
\newcommand{\ee}{\end{equation}}
\def\reff#1{(\protect\ref{#1})}

\usepackage{dsfont}
\def\reff#1{(\protect\ref{#1})}

\begin{document}

\title{Exact value for the average optimal cost of bipartite traveling-salesman and 2-factor\\   problems in two dimensions}
\author{Riccardo Capelli}\email{r.capelli@fz-juelich.de}
\affiliation{Dipartimento di Fisica, University of Milan and INFN, via Celoria 16, 20133 Milan, Italy }
\affiliation{Computational Biomedicine Section, Institute of Advanced Simulation IAS-5 and Institute of Neuroscience and Medicine INM-9, Forschungszentrum J\"ulich, Wilhelm-Johnen-Stra\ss{}e, 52425 J\"ulich, Germany}
\author{Sergio Caracciolo}\email{sergio.caracciolo@mi.infn.it}
\affiliation{Dipartimento di Fisica, University of Milan and INFN, via Celoria 16, 20133 Milan, Italy}
\author{Andrea Di Gioacchino}\email{andrea.digioacchino@unimi.it}
\affiliation{Dipartimento di Fisica, University of Milan and INFN, via Celoria 16, 20133 Milan, Italy}
\author{Enrico M. Malatesta}\email{enrico.m.malatesta@gmail.com}
\affiliation{Dipartimento di Fisica, University of Milan and INFN, via Celoria 16, 20133 Milan, Italy}

\date{\today}
\begin{abstract}
We show that the average optimal cost for the traveling-salesman problem in two dimensions, which is the archetypal problem in combinatorial optimization, in the bipartite case, is simply related to the average optimal cost of the assignment problem with the same Euclidean, increasing, convex weights. In this way we extend a result already known in one dimension where exact solutions are avalaible. The recently determined average optimal cost for the assignment when the cost function is the square of the distance between the points provides therefore an exact prediction
$$
\overline{E_N} = \frac{1}{\pi}\, \log N
$$ 
for large number of points $2N$. As a byproduct of our analysis also the loop covering problem has the same optimal average cost.
We also explain why this result cannot be extended at higher dimensions. We numerically check the exact predictions.
\end{abstract}
\maketitle

\section{Introduction}

The Traveling Salesman Problem (TSP)~\cite{Karp, Frieze} can be formulated in few words: what is the shortest tour which goes through $N$ given points? But this is well known to be a computationally intractable problem. The number of possible solutions increases exponentially with the number of points and there is not a known algorithm able to find the solution in a time that increases less than exponentially with $N$.

When the emphasis is shifted from the research of the solution (in the worst case) to the typical properties of the solution in a class of possible instances, the statistical properties of the optimal solutions can be described by a zero-temperature statistical model. This approach has been tremendously fruitul~\cite{Kirkpatrick1983, Vannimenus1984, fu1986application, Sourlas1986, mezard1987spin, mezard2009information}.
The random model in which the distances between the cities are independent and equally distributed random variables has been deeply studied~\cite{Orland1985, Mezard1986a, Mezard1986, Krauth1989}.
Much less is known for the Euclidean version of the random problem, where the position of the points are chosen at random in a finite domain of $\mathbb{R}^d$, so that the distances of the points are now correlated~\cite{BHH, Percus1996, Cerf1997}. Of course, for large $d$, the effects of these correlations are smaller and smaller and the methods used to deal with the problem in absence of correlations becomes more and more effective. 

In the Euclidean version of the problem, we associate to the step in the tour from the $i$-th point with coordinate $x_i$ to the $j$-th point with coordinate $x_j$ a cost 
\be
c_p(x_i, x_j) = c_p(x_i - x_j) := \| x_i - x_j\|^p \, ,\label{c}
\ee
with $p\in \mathbb{R}$  and $\| x_i - x_j \|$ the Euclidean distance between the two points. In the bipartite version of the problem the set of $2N$ points is partitioned in two subsets each with $N$ points and steps are allowed only from points in one subset to points in the other subset, in the monopartite version all the points can be reached from any other point.
Interestingly enough in $d=1$, when $p>1$, that is when the cost function is convex and increasing, the search for the optimal tour can be exactly solved both in the bipartite~\cite{Caracciolo:171}, as well as in the monopartite~\cite{Caracciolo:175}, version of the problem.

There have been, recently, what we consider three relevant progresses in the field:
\begin{itemize}
\item[i)] for other optimization problems similar to the TSP, the monopartite and bipartite versions have different optimal cost properties. For example for the matching, 1-factor and 2-factor (or loop-covering) problems, the optimal cost is expected to be a self-averaging quantity whose average scales according to
\be
\overline{E_N^{(p,d)}} \sim N^{1- \frac{p}{d}}
\ee
(see~~\cite{BHH} for a proof in the case $p=1$). On the other hand, in the bipartite version~\cite{Ajtai, Caracciolo:158, Caracciolo:159} it is expected that
\be
\overline{E_N^{(p,d)}}  \sim 
\begin{cases}
N^{1- \frac{p}{2}} & \hbox{for \, } d=1\\
N^{1- \frac{p}{2}}  (\log N)^\frac{p}{2} & \hbox{for \, } d=2 \\
N^{1- \frac{p}{d}} & \hbox{for \, } d>2
\end{cases}
\ee
that is a larger average cost with respect to the monopartite case when $d\leq 2$. Moreover, in the bipartite case the optimal cost is expected to be not self-averaging;
\item[ii)] in the bipartite case it is always true~\cite{Caracciolo:2m} that the total optimal cost of the TSP $E_\mathcal{H}^*$ is larger than the total optimal cost of the 2-factor problem $E_{\mathcal{M}_2}^*$, which is larger than twice the total optimal cost of the corresponding matching problem (assignment) $E_{\mathcal{M}_1}^*$
\begin{equation}
	E_\mathcal{H}^*\geq E_{\mathcal{M}_2}^* \geq 2 E_{\mathcal{M}_1}^* \, .\label{ine}
\end{equation}
In~\cite{Caracciolo:171} it has been shown that in $d=1$, in the asymptotic limit of an infinitely large number points, this bound is saturated, that is the total optimal cost of the TSP, rescaled with $N^{1- \frac{p}{2}}$, is exactly twice the total rescaled optimal cost of the assignment problem, and, therefore, all the three quantities coincide.
\item[iii)] in the bipartite case, in $d=2$ and $p=2$, thanks to a deep connection with the continuum version of the problem, that is the well known {\em transport problem}, it has been possible to compute, exactly,  the total optimal cost of the assignment problem in the asymptotic limit of an infinitely large number points ~\cite{Caracciolo:158, Caracciolo:162, Caracciolo:163, Ambrosio2016}:
\be
\overline{E_N^{(2,2)}} = \frac{1}{2\pi} \, \log N\, . \label{mat}
\ee
\end{itemize}
We considered, therefore, the possibility that also in $d=2$, and $p>1$, exactly, thanks to the logarithmic violation present in the bipartite case, the asymptotic total cost of the TSP can be exactly twice the one of the assignment, that for $p=2$ is also exactly known. Indeed, this is the case!

The paper is organized as follows. In Section~\ref{model} we present the model. 
In Section~\ref{scaling} we present an argument to justify our strategy.
In Section~\ref{numerics} we provide evidence by numerical simulations how our result is established for large number of points. We also examined the  case $p=1$, which is the most largely considered in the literature.

\section{The model}\label{model}
Consider a generic graph $\mathcal{G}=(V,E)$ where $V$ is its set of vertices and $E\subset V^2$ its set of edges.
Let $w_e>0$ be a weight associated to the edge $e\in E$. We shall consider the complete (monopartite) graph $\mathcal{K}_N$, whose vertex set has cardinality $N$ and $E=V^2$ and the complete bipartite graph $\mathcal{K}_{N,N}$, whose vertex set is $V=V_1 \cup V_2$ with $V_1$ and $V_2$ disjoint sets of cardinality $N$,  and $E= V_1\times V_2$.

\begin{figure}
	\centering
	\includegraphics[width=0.7\columnwidth,keepaspectratio]{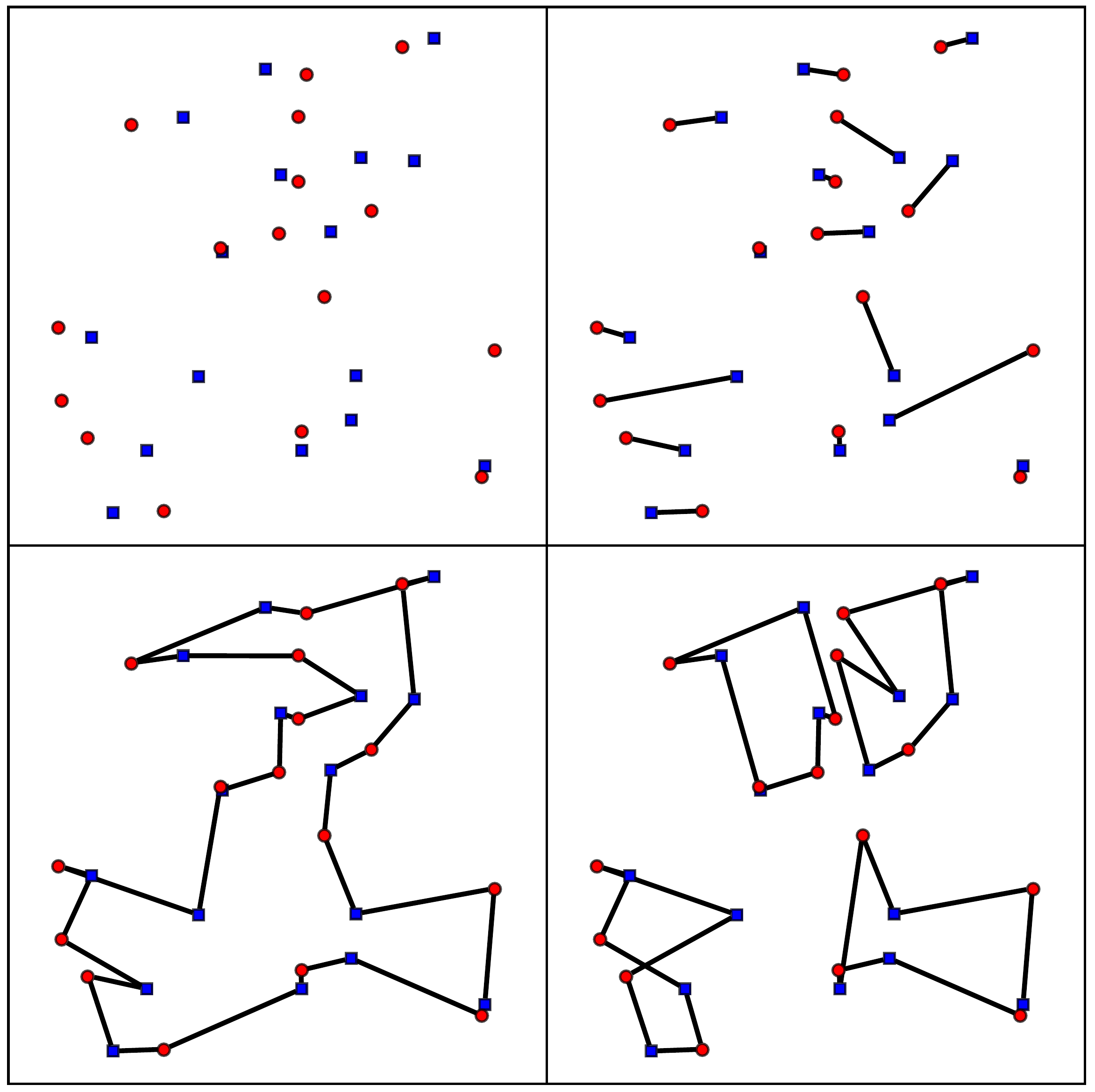}
	\caption{On the same instance with $N=16$ blue (squared) and red (disk) points (top left panel), we draw an arbitrarily-chosen example of each class of spanning subgraph we are considering: a 1-factor (top right), an Hamiltonian cycle (bottom left) and a 2-factor (bottom right). }\label{all}
\end{figure}

Let $\mu =(V,E')$ be a spanning subgraph of $\mathcal{G}$, that is $E'\subset E$. We can define a total cost associated to $\mu$ according to
\be
E[\mu] = \sum_{e\in \mu} w_e\, .
\ee
We shall consider three different classes of spanning subgraphs~$\mathcal{M}$. 
The set $\mathcal{M}_1$ of 1-factor (matching), where each vertex belongs to one and only one edge, the set $\mathcal{M}_2$ of 2-factor (2-matching or loop covering), where each vertex belongs to two edges, and the set $\mathcal{H}$ of Hamiltonian cycles, that is 2-factor formed by only one cycle (see Fig. \ref{all}). 
The assignment, respectively 2-factor, TSP, problems amounts to the search of the subgraph $\mu^*$ in $\mathcal{M}_1$, respectively $\mathcal{M}_2$, $\mathcal{H}$, which is optimal, in the sense of minimal total cost
\be
E_\mathcal{M}^* = E_\mathcal{M}[\mu^*] = \min_{\mu\in \mathcal{M}} E[\mu] \, 
\ee
with $\mathcal{M}$ respectively $\mathcal{M}_1, \mathcal{M}_2, \mathcal{H}$.
In the Euclidean version of our bipartite optimization problems, we consider the immersion of $\mathcal{K}_{N,N}$ in an open subset $\Omega \subset \mathbb{R}^d$. $V_1$, respectively $V_2$, will be identified by the set of $N$ points with coordinates $r_i$'s, that we shall call the red points, respectively by the set of $N$ points with coordinates $b_j$'s, that we shall call the blue points. 
 Let  $(i, j)$ be the edge connecting the $i-$th red vertex with the $j-$th blue vertex. We give a weight  $w_{ij} = c_p(r_i - b_j)$, with $p \geq 1$ as in Eq.~\reff{c}. 
 
 Of course, in the monopartite Euclidean version there is only one set of points.
 
 In the random Euclidean version of the problem, each possible instance is obtained by choosing at random, with a given law, the position of the points in $\Omega$. For example we shall consider $\Omega=[0,1]^d$ and the flat distribution.
 We denote by $\overline{E^*}$ the average, over all instances, of the optimal total cost $E^*$.

\section{Scaling argument}\label{scaling}
%
In this section we will provide a scaling argument to support our claim, that is,  also in two dimensions, for any given choice of the positions of the points, in the asymptotic limit of large $N$,  the cost of the bipartite TSP converges to twice the cost of the assignment. 

Given an instance, let us consider the optimal assignment $\mu^*$ on them. Let us now consider $N$ points which are taken between the red an blue point of each edge in $\mu^*$ and call $\mathcal{T}^*$ the optimal \emph{monopartite} TSP solution on these points. For simplicity, as these $N$ points we take the blue points. 

We shall use $\mathcal{T}^*$ to provide an ordering among the red and blue points. Given two consecutive points in $\mathcal{T}^*$, for example $b_1$ and $b_2$, let us denote by $(r_1,b_1)$ and $(r_2,b_2)$ the two edges in $\mu^*$ involving the blue points $b_1$ and $b_2$ and let us consider also the new edge $(r_1, b_2)$.
We know that, in the asymptotic limit of large $N$, the typical distance between two matched points in $\mu^*$ scales as $ (\log N/N)^{1/2}$ while the typical distance between two points matched in the monopartite case scales only as $1/N^{1/2}$, that is (for all points but a fraction which goes to zero with $N$)
\begin{equation}
\begin{split}
& w_{(b_1,r_1)} = \left( \alpha_{11} \frac{\log N}{N} \right)^{\frac{p}{2}}, \\ 
& w_{(b_2,r_1)} =\left[\beta_{22} \frac{1}{N} + \alpha_{11} \frac{\log N}{N} - \gamma \frac{\sqrt{\log N}}{N} \right]^{\frac{p}{2}}.
\end{split}
\end{equation}
where $(\alpha_{11} \log N / N)^{1/2}$ is the length of the edge $(r_1,b_1)$ of $\mu^*$, $(\beta_{22} / N)^{1/2}$ is the length of the edge $(b_1, b_2)$ of $\mathcal{T}^*$ and $\gamma = 2 \sqrt{\alpha_{11} \beta_{22}} \cos\theta$, where $\theta$ is the angle between the edges $(r_1,b_1)$ of $\mu^*$ and $(b_1,b_2)$ of $\mathcal{T}^*$.

\begin{figure}
	\includegraphics[]{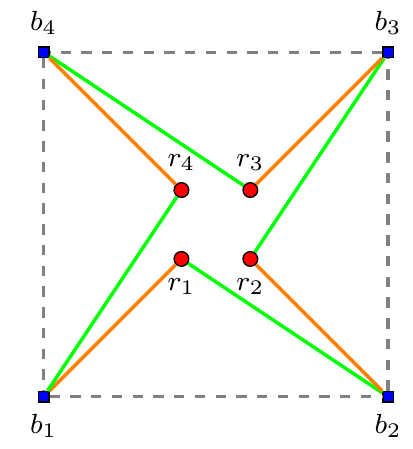}
	\caption{The optimal assignment $\mu^*$ is given by the orange edges $\{ (r_1, b_1), (r_2 ,b_2), (r_3, b_3), (r_4 ,b_4) \}$. 
	The monopartite TSP (gray dashed edges) among blue points provides the necessary ordering. In order to obtain the TSP ${b_1, r_1, b_2, r_2, b_3, r_3, b_4, r_4, b_1}$ in the bipartite graph we have to add the green edges $\{((r_1, b_2), (r_2, b_3), (r_3, b_4), (r_4, b_1)\}$.
	}\label{figscale}
\end{figure}

This means that, typically,  the difference in cost 
\be
\Delta E = w_{(b_2,r_1)} - w_{(b_1,r_1)} \sim \frac{(\log N)^{\frac{p-1}{2}}}{N^{\frac{p}{2}}}
\ee
is small as compared to the typical cost $(\log N/N)^\frac{p}{2}$ of one edge in the bipartite case.
To obtain a valid TSP solution, which we call $h^A$, we add to the edges $\mu^* = \{(r_1,b_1), \dots,(r_N, b_N)\}$ the edges $\{(r_1,b_2), \dots, (r_{N-1}, b_N), (r_N, b_1)\}$, see Figure~\ref{figscale}.

Of course $h^A$ is not, in general, the optimal solution of the TSP. However, because of Eq.~\reff{ine}, we have that
\be
E_\mathcal{H}[h^A] \geq E_\mathcal{H}^*  \geq E_{\mathcal{M}_2}^* \geq 2\, E_{\mathcal{M}_1}^*
\ee
and we have shown that, for large $N$, $E_\mathcal{H}[h^A]$ goes to $2\, E_{\mathcal{M}_1}^*$ and therefore also $E_\mathcal{H}^*$ must behave in the same way. As a byproduct of our analysis also $E_{\mathcal{M}_1}^*$ for the loop covering problem has the same optimal average cost. Note also that our argument is purely local and therefore it does not depend in any way on the type of boundary conditions adopted. Since in the case of periodic boundary conditions, as shown in~\cite{Caracciolo:163}, it holds~(\ref{mat}), we get that the average optimal cost of both the TSP and 2-factor goes for large $N$ to 2 times the optimal assignment.


Notice that an analogous construction can be used in any number of dimensions. However, the success of the procedure lies in the fact that the typical distance between two points in $\mu^*$ goes to zero slower than the typical distance between two consecutive points in the monopartite TSP. This is true only in one and two dimensions, and it is related to the importance of fluctuations in the number of points of different kinds in a small volume.

This approach allowed us to find also an approximated solution of the TSP which improves as $N\to\infty$. However, this approximation requires the solution of a \emph{monopartite} TSP on $N/2$ points, corroborating the fact that the bipartite TSP is a hard problem (from the point of view of complexity theory). 

A similar construction can be used to achieve an approximated solution also for the 2-factor problem. In this case, instead of solving the monopartite TSP on a point chosen within each edge of $\mu^*$, one should solve the monopartite matching problem on this set of points, obtaining a matching $\mathcal{M}^*$. 
Once more let  us denote by $(r_1,b_1)$ and $(r_2,b_2)$ the two edges in $\mu^*$ which give rise to two matched points in $\mathcal{M}^*$,  and collect  them together with the  edges $(r_1, b_2)$ and $(r_2, b_1)$.
Repeating the above procedure for each couple of points matched in $\mathcal{M}^*$, the union of the edges obtained  gives a valid 2-factor whose cost tends, in the limit of large $N$,  to twice the cost of the optimal assignment in one and two dimensions. Notice that, in this case, the procedure is much more efficient because the solution of the matching problem is polynomial in time.
\begin{figure*}[ht]
	\begin{subfigure}[t]{0.49\linewidth}
		\centering
		\includegraphics[width=0.95\columnwidth]{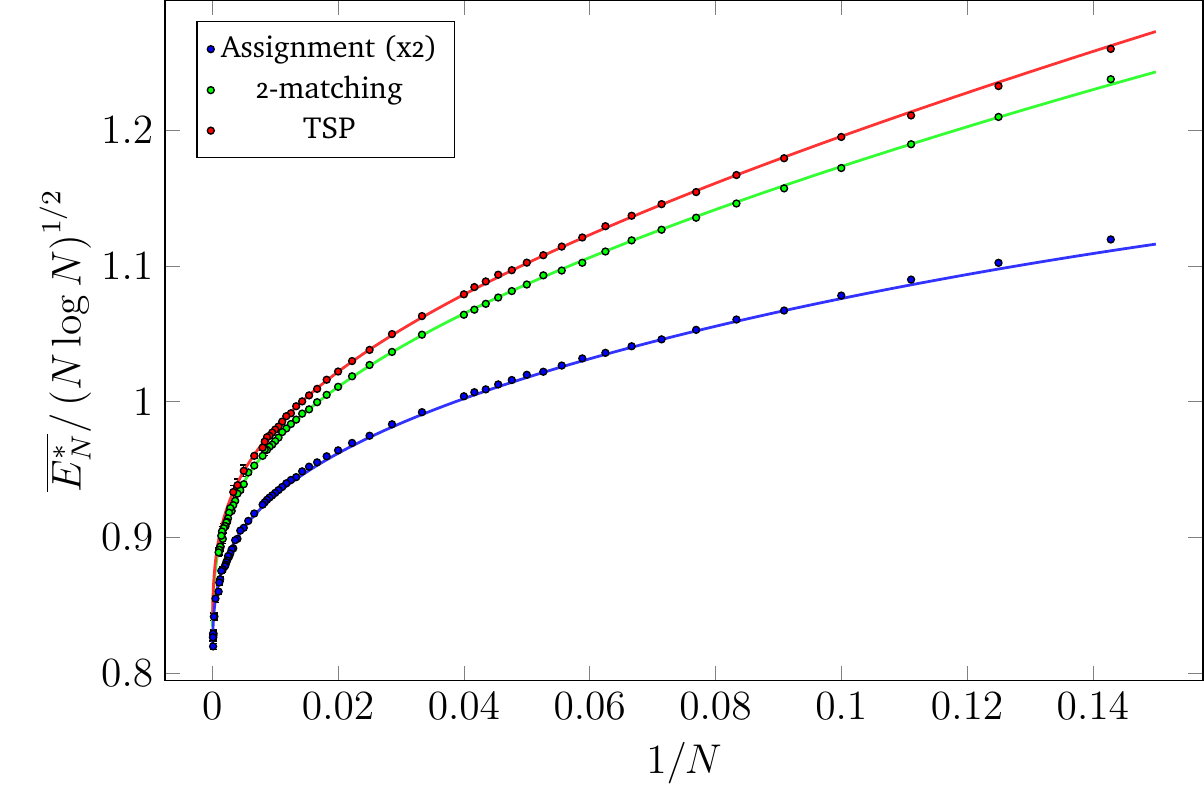}
		\label{Fig::p=1}
	\end{subfigure} \hfill
	\begin{subfigure}[t]{0.49\linewidth}
		\centering
		\includegraphics[width=0.95\columnwidth]{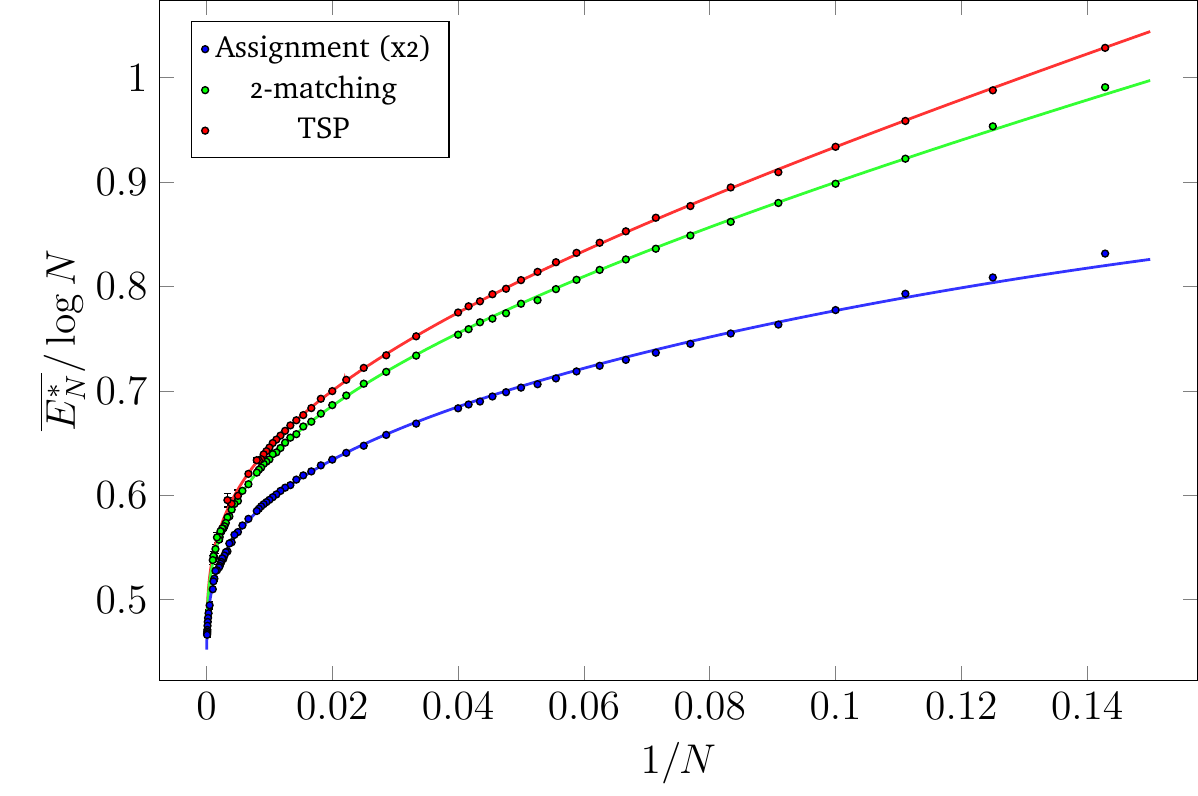}
		\label{Fig::p=2}
	\end{subfigure}
	\caption{Numerical results for $p=1$ (left panel) and $p=2$ (right panel) for the TSP (red points, top), the 2-factor (green points, middle) and 2 times the assignment problem (blue points, bottom) in the open boundary condition case. Continuous lines are numerical fit to the data.}
	\label{Fig::AverageCost}
\end{figure*}
\begin{table*}[ht]
	\centering
	\begin{tabular}{llll} 
		{$p=1$} & $a_{1}$ & $a_{2}$ & $a_{3}$ \\
		\hline
		TSP & 0.717(2) & 1.32(1) & $-0.513(1)$ \\ 
		2-factor & 0.714(2) & 1.31(1) & $-0.58(2)$ \\ 
		Assignment & 0.714(2) & 1.17(2) & $-0.77(2)$ \\  
	\end{tabular}
	\quad \quad \quad \quad \quad \quad \quad \quad
	\begin{tabular}{llll} 
		{$p=2$} & $a_{1}$ & $a_{2}$ & $a_{3}$ \\
		\hline
		TSP & 0.321(5) & 1.603(2) & $-0.428(6)$ \\ 
		2-factor & 0.319(4) & 1.577(2) & $-0.547(7)$ \\ 
		Assignment & 0.31831 & 1.502(2) & $-1.05(1)$  \\ 
	\end{tabular}
	\caption{Comparison between fit factors in assignment and TSP, for $p=1$, $2$.  We have doubled the factors for the assignment to verify our hypothesis. For $p=2$, we have reported the theoretical value of $a_1$ which is $1/\pi$. 
	}
	\label{tab:p1}
\end{table*}

\section{Numerical Results}\label{numerics}

We have confirmed our theoretical predictions performing numerical simulations on all the three models previously presented: assignment, bipartite 2-factor, and bipartite TSP. We have considered the case of open boundary conditions.

For what concerns the assignment problem, many polynomial-time algorithms are available in the literature, as the famous Hungarian algorithm~\cite{kuhn1955hungarian}. We have implemented an in-house assignment solver based on the LEMON optimization library~\cite{dezso2011lemon}, which is based on the Edmonds' blossom algorithm~\cite{edmonds1965paths}. In the case of the 2-factor and TSP, the most efficient way to tackle numerically those problems is to exploit their \textit{linear} or \textit{integer programming} formulation. 

To validate our argument, we solved for both assignment and 2-factor problem (with $p=1,2$), $10^{5}$ independent instances for $2 \leq N \leq 125$, $10^{4}$ independent instances for $150 \leq N \leq 500$, and $10^{3}$ independent instances for $600 \leq N \leq 1000$. In the TSP case, the computational cost is dramatically larger; for this reason the maximum number of points we were able to achieve with a good numerical precision using integer programming was $N=300$, also reducing the total number of instances. 

An estimate of the asymptotic average optimal cost and finite size corrections has been obtained using the fitting function for $p=1$
\begin{equation}
\label{eq:p1_cost}
f^{(p=1)}(N) = \sqrt{N\log{N}} \left( a_{1}  +  \frac{a_{2}}{\log N} + \frac{a_{3}}{\log^2 N} \right)
\end{equation}
while, for $p=2$
\begin{equation}
\label{eq:p2_cost}
f^{(p=2)}(N) =  \log N \, \left( a_{1}  +  \frac{a_{2}}{\log N} + \frac{a_{3}}{\log^2 N} \right) \,.
\end{equation}
These are the first 3 terms of the asymptotic behavior of the cost of the  assignment problem~\cite{Ajtai,Caracciolo:158}. 
Parameters $a_2$ and $a_3$ for $p=2$ were obtained fixing $a_1$ to $1/\pi$. In Figure~\ref{Fig::AverageCost} we plot the data and fit in the case of open boundary conditions. Results are reported in Table~\ref{tab:p1}.

To better confirm the behavior of the average optimal cost of the TSP, we also performed some numerical simulations using a much more efficient solver, that is the Concorde TSP solver~\cite{applegate2006concorde}, which is based on an implementation of the Branch-and-cut algorithm proposed by Padberg and Rinaldi~\cite{padberg1991branch}. The results for the leading term of the asymptotic average optimal cost are confirmed while a small systematic error due to the integer implementation of the solver is observed in the finite size corrections.

\section{Conclusions}\label{fine}
In this work we have considered three combinatorial optimization problems, the matching problem, 2-factor problem and TSP, where the cost is a convex increasing function of the point distances. 
Previous investigations have been performed in the one-dimensional case, by means of exact solutions~\cite{Caracciolo:171}.
Here we analyzed the bipartite version of these problems in two dimensions, showing that, as already obtained in one dimension:
\begin{equation}
	\lim_{N\to\infty} \frac{\overline{E_\mathcal{H}^*}}{\overline{E_{\mathcal{M}_1}^*}} = \lim_{N\to\infty} \frac{\overline{E_{\mathcal{M}_2}^*}}{\overline{E_{\mathcal{M}_1}^*}} = 2  \, .
\end{equation}
This implies, for the special case $p=2$,  by using~\reff{mat}, our main exact result, that is $\lim_{N\to\infty} (\overline{E^*_\mathcal{H}}/ \log N) =   1/\pi$.
In general, the evaluation of  $\overline{E^*_\mathcal{H}}$ and $\overline{E^*_{\mathcal{M}_2}}$ for large $N$ is reduced to the solution of the matching problem which requires only polynomial time.
This seems to be a peculiar feature of the bipartite problem: the monopartite TSP cannot be approached in a similar way.
As a byproduct of our analysis, we provided in Sec. \ref{scaling} two approximate algorithms, for the bipartite TSP and the bipartite 2-factor: both are guaranteed to give a solution with optimal cost for large $N$. The first algorithm allows to solve the bipartite TSP on $N$ points solving the monopartite TSP with $N$ points (notice that, on principle, the bipartite version consists of $2N$ points). The second allows to exploit the fast Hungarian algorithm to obtain an approximate solution of the 2-factor problem.

\section*{Acknowledgements}
This project has received funding from the European Union's Horizon 2020 Research and Innovation Programme under Grant Agreement No. 785907 (HBP SGA2).
We thank Gabriele Sicuro for sharing with us his numerical results on the matching.

\bibliographystyle{unsrt}
\bibliography{AssignmentANDTsp}

\end{document}